# Radial basis function network using Lambert-Tsallis $W_q$ function

J. L. M. da Silva       F. V. Mendes       R. V. Ramos

Jorgemouta95@gmail.com   fernandovm@gmail.com   rubens.ramos@ufc.br

*Lab. of Quantum Information Technology, Department of Teleinformatic Engineering – Federal University of Ceara - DETI/UFC, C.P. 6007 – Campus do Pici - 60455-970 Fortaleza-Ce, Brazil.*

*Abstract*

The present work brings two applications of the Lambert-Tsallis $W_q$ function in radial basis function networks (RBFN). Initially, a RBFN is used to discriminate between entangled and disentangled bipartite of qubit states. The kernel used is based on the Lambert-Tsallis $W_q$ function for $q = 2$ and the quantum relative disentropy is used as distance measure between quantum states. Following, a RBFN with the same kernel is used to estimate the probability density function of a set of data samples.

*Key words* – Lambert-Tsallis $W_q$ function; Quantum relative disentropy; Radial basis function network;

## 1. Introduction

The Lambert *W* function has several applications in physics and engineering [1-3]. Its generalization, the recently proposed Lambert-Tsallis $W_q$ function [4] was used to define the disentropy [4] that, for example, has several applications in quantum and classical information theory [5]. In the present work we show that $W_q$ can be used to construct a useful radial basis function that can be successfully used in a radial basis function network (RBFN). Two RBFN were constructed. The first one is a classifier trained to identify entangled and disentangled bipartite of qubit states. It uses the quantum relative disentropy [5] as distance measure between quantum states. The Wootters' concurrence [6] is used to measure the error rate of the RBFN. Following, a second RBFN is used to estimate the probability density function of a set of data samples. The test of this RBFN was done considering two situations: data samples from normal and Cauchy distributions.

This work is outlined as follows: In Section 2, a brief review of RBFN, Lambert-Tsallis function and disentropy is provided; In Section 3 the proposed classifier is introduced and the numerical results are shown. In Section 4 the proposed PDF estimator is presented and the numerical results are shown. At last, the conclusions are drawn in Section 5.

## 2. Radial basis function network, Lambert-Tsallis $W_q$ function and disentropy

Radial basis function network can successfully be used to realize function approximation and pattern recognition, for example. Its basic structure with only one hidden layer is shown in Fig. 1.

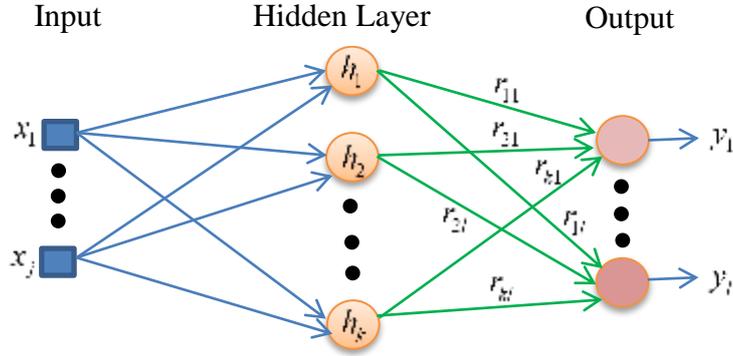

Fig. 1 – RBFN with one hidden layer.

The *t*-th output, $y_t$, is given by

$$y_t(\vec{x}) = \sum_{n=1}^{k} r_{nt} h_n \left[ d(\vec{x}, \vec{x}_n^c); \delta_n; c_n \right]. \tag{1}$$

In (1), the input variable $\vec{x}$ is a function of the input data $\{x_1,\ldots,x_j\}$, $h_n$ is the radial basis function of the *n*-th neuron. It has three parameters: $\vec{x}_n^c$ (the center), $\delta_n$ (the width) and $c_n$ (the type of RBF used). Basically, the value of $h_n$ is maximum when the distance given by the function $d(\cdot)$ between its center and the input variable is zero and it decreases when the distance increases. Functions like Gaussian, *q*-Gaussian and Cauchy function have been used as RBF [7,8]. The distance function $d$ can be, for example, an Euclidean distance or an entropic distance. A RBFN can use different RBFs, in this case the parameter $c_n$ identifies the type of RBF used. At last, $r_{nt}$ are the coefficients found during the training stage. The training consists of finding out the best values of $\vec{x}_n^c$, $\delta_n$, $c_n$, $r_{nt}$ and the number of neurons in the hidden layer, such that the error during the training using known examples is minimal. The size and quality of the training set is also important for a good performance of the classifier. It is common to use a genetic algorithm or another heuristic to train the RBFN.

The RBF proposed in this work is

$$h\left[d(x,x_c);\delta;q\right] = \frac{C_1}{1+W_q\left[\delta d(x,x_c)\right]} - C_2, \tag{2}$$

where $C_1$ and $C_2$ are parameters used to normalize the function $h$ and $W_q$ is the Lambert-Tsallis function that was introduced in [4]. Basically, it is the function the solves the equation

$$W_q(z) e_q^{W_q(z)} = z. \tag{3}$$

In (3) $q$ is the Tsallis non-extensivity parameter [9] and $e_q(x)$ is the $q$-exponential. When $q = 1$, one has $e_q(x) = e^x$ and, hence, $W_{q=1}$ is the famous Lambert $W$ function. In Fig. 2 it is shown the plot of eq. (2) for two cases: I) $C_1 = 1$, $C_2 = 0$, $q = 1$, $\delta = 1$, $d = (x\text{-}50)^2$. II) $C_1 = 2$, $C_2 = 1$, $q = 2$, $\delta = 1$, $d = (x\text{-}50)^2$.

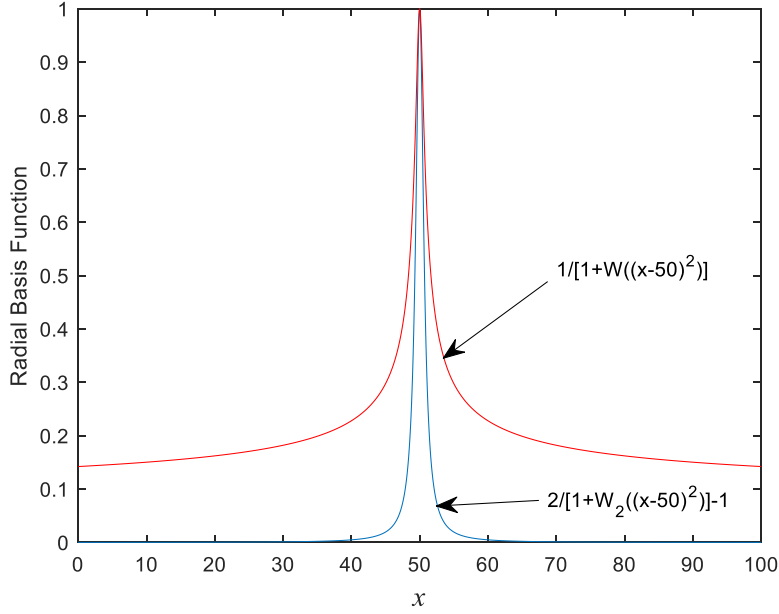

Fig. 2 – Radial basis functions based on $W$ and $W_2$.

The quantum disentropy is defined as [4]

$$D_q(\rho) = \sum_n \lambda_n^q W_q(\lambda_n), \tag{4}$$

where $\lambda_i$ is the $i$-th eigenvalue of the density matrix $\rho$. The quantum relative disentropy, by its turn, is given by [5]

$$D_q^R(\rho\|\Gamma) = \sum_n \lambda_n^q \left| W_q(\lambda_n) - W_q(\gamma_n) \right|. \tag{5}$$

where $\lambda_n$ and $\gamma_n$ are, respectively, the eigenvalues of the density matrices $\rho$ and $\Gamma$.

## 3. Classifier of $C_2 \otimes C_2$ quantum states using RBFN with Lambert-Tsallis Function

Since the goal is to show that $W_q$ can be used to construct a useful radial basis function, initially we chose to implement a classifier able to discriminate between entangled and disentangled two-qubit states. This problem is easily solved using Wootter's concurrence [6], hence we used it in order to measure the error rate of the RBFN implemented. The classifier proposed has one input (the input quantum state), one hidden layer and a single output value. It is shown in Fig. 3.

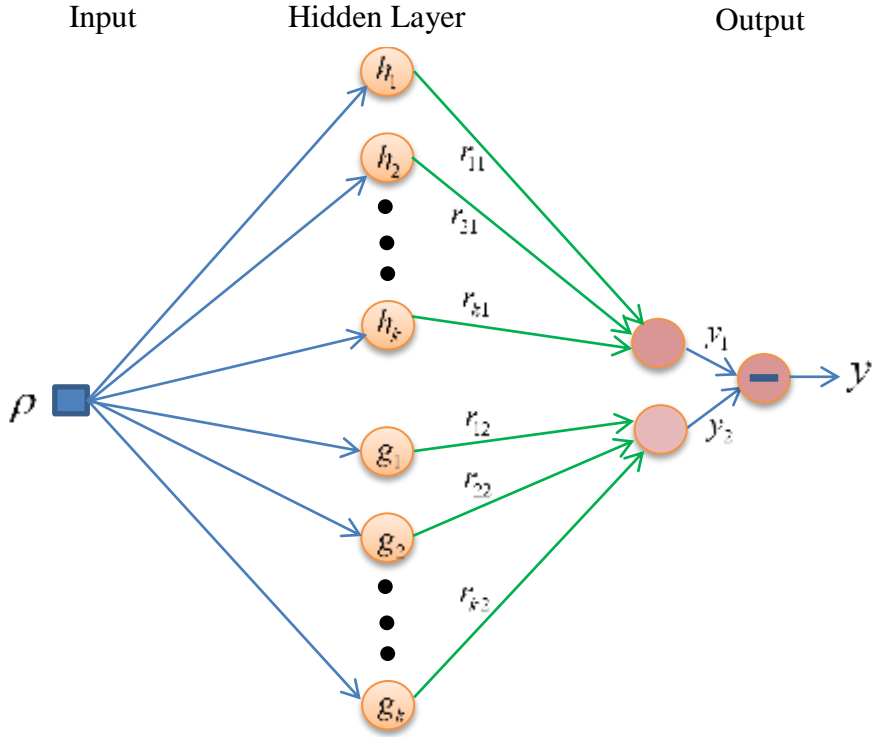

Fig. 3 – RBFN used to discriminate between entangled and disentangled two-qubit states.

The distance function used is the quantum relative disentropy,

$$d(\rho, \Gamma_n^c) = D_q^R(\rho \| \Gamma_n^c). \tag{6}$$

In (6) $\Gamma_n^c$ are the 'central' quantum states. Using (6) and (2) in (1) one gets the equation that describes the proposed classifier

$$y_1(\rho) = \sum_{n=1}^{k} \frac{r_{n1}}{1 + W_q\left[\delta_{n1} D_q^2(\rho \| \Gamma_n^c)\right]} \tag{7}$$

$$y_2(\rho) = \sum_{n=1}^{k} \frac{r_{n2}}{1 + W_q\left[\delta_{n2} D_q^2(\rho \| \Phi_n^c)\right]} \tag{8}$$

$$y(\rho) = y_1(\rho) - y_2(\rho). \tag{9}$$

The central states $\Gamma_n^c$ are entangled while $\Phi_n^c$ are disentangled states. The input state $\rho$ is considered entangled (disentangled) by the RBFN if $y(\rho) > 0$ ($y(\rho) \leq 0$). The heuristic used to training the RBFN (to find the values of $\Gamma_n^c$, $\Phi_n^c$, $r_{n1}$, $r_{n2}$, $\delta_{n1}$ and $\delta_{n2}$ that minimizes the error rate) was the *Differential Evolution* (*DE*), since the solutions are directly coded in real values instead of binary strings. We used $q = 2$ and 20 neurons in the hidden layer, $k = 10$ in (7)-(8). The initial training set was composed by 2500 entangled states and 2500 disentangled states chosen randomly. The *DE* algorithm used a population with 20 individuals and crossover and mutation rates equal to 0.75 and 0.25, respectively. The training took 1,000 generations. After the training, we used a test set with 1,000,000 states (500,000 entangled and 500,000 disentangled) chosen randomly and the classifier discriminated them correctly in 89.42% of the cases. The correct discrimination is more complicated when the input is an entangled state with a very low value of entanglement. In order to check the performance of the classifier in this region, we considered the scenarios described in the tables I and II:

Table I – Training sets. *C* is the Wootter's concurrence.

| Training set – 5,000 states randomly chosen ||
|---|---|
| $Str_0$ | 2,500 disentangled states and 2,500 entangled states without any restriction. |
| $Str_1$ | 2,500 disentangled states and 2,500 entangled states with $C \leq 0.1$. |
| $Str_2$ | 2,500 disentangled states and 2,500 entangled states with: $C \leq 0.1$ (75%) and $0.1 < C \leq 0.2$ (25%). |
| $Str_3$ | 2,500 disentangled states and 2,500 entangled states with: $C \leq 0.1$ (50%) and $0.1 < C \leq 0.2$ (50%). |
| $Str_4$ | 2,500 disentangled states and 2,500 entangled states with: $C \leq 0.1$ (25%) and $0.1 < C \leq 0.2$ (75%). |
| $Str_5$ | 2,500 disentangled states and 2,500 entangled states with: $C \leq 0.1$ (25%); $0.1 < C \leq 0.2$ (25%); $0.2 < C \leq 0.3$ (25%); $0.3 < C \leq 0.4$ (25%). |

Table II – Test sets.

| Test set – 1,000,000 states randomly chosen ||
|---|---|
| $Stst_1$ | Entangled and disentangled states without any restriction |
| $Stst_2$ | Only disentangled states |
| $Stst_3$ | Only entangled states |
| $Stst_4$ | Only entangled states with $C \leq 0.1$ |
| $Stst_5$ | Only entangled states with $0.1 < C \leq 0.2$ |
| $Stst_6$ | Only entangled states with $0.2 < C \leq 0.3$ |
| $Stst_7$ | Only entangled states with $0.3 < C \leq 0.4$ |
| $Stst_8$ | Only entangled states with $0.4 < C \leq 0.5$ |
| $Stst_9$ | Only entangled states with $0.5 < C \leq 0.6$ |

The success rates for the different scenarios are shown in Table III.

Table III – Success rate when the training and test sets used are $Str_i$ and $Stst_j$.

| $q = 2$ | $Stst_1$ | $Stst_2$ | $Stst_3$ | $Stst_4$ | $Stst_5$ | $Stst_6$ | $Stst_7$ | $Stst_8$ | $Stst_9$ |
|---|---|---|---|---|---|---|---|---|---|
| $Str_0$ | 87,00% | 85,31% | 86,93% | 75,56% | 99,65% | 100,00% | 100,00% | 100,00% | 100,00% |
| $Str_1$ | **89,42%** | 75,69% | **89,34%** | **87,80%** | **100,00%** | 100,00% | 100,00% | 100,00% | 100,00% |
| $Str_2$ | 87,60% | 83,69% | 87,51% | 78,06% | 99,99% | 100,00% | 100,00% | 100,00% | 100,00% |
| $Str_3$ | 86,06% | 86,07% | 85,98% | 72,62% | 99,92% | 100,00% | 100,00% | 100,00% | 100,00% |
| $Str_4$ | 84,01% | 89,08% | 84,00% | 65,80% | 99,50% | 100,00% | 100,00% | 100,00% | 100,00% |
| $Str_5$ | 81,86% | **91,76%** | 81,82% | 59,45% | 97,70% | 100,00% | 100,00% | 100,00% | 100,00% |

As it can be noted in Table III, entangled states with concurrence larger than 0.2 are always correctly discriminated by the RBFN.

## 4. Probability density function estimator usand RBFN and $W_2$

Sometimes one is interested in the determination of the probability density function of a physical process, like in quantum state tomography [10]. In this cases only the data sample whose probability density function is not known in advance are availiable. Hence, one has to estimate the PDF using the avaliable set of data. There are several PDF estimators in the literature [11]. The easiest way to estimate the PDF from avaliable data is to construct the histogram. However, this estimation is coarse. Here, we propose a RBFN for PDF estimation using $W_2$ in the kernel. The proposed RBFN is shown in Fig. 4.

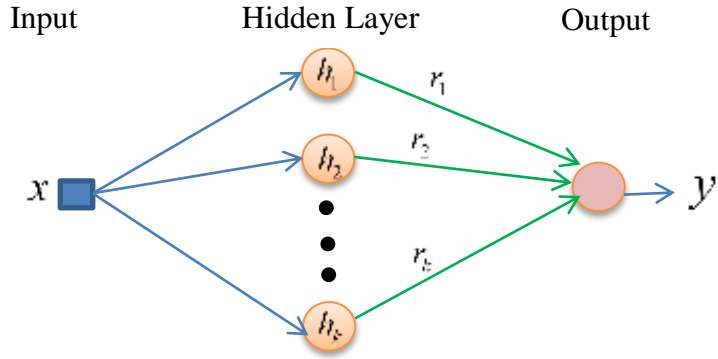

Fig. 4 – RBFN for PDF estimation.

The RBFN output is

$$y(x) = \sum_{n=1}^{k} r_n \left[ \frac{2}{1 + W_2 \left[ \delta_n (x - x_n^c)^2 \right]} - \frac{1}{2} \right]. \tag{10}$$

In order to get the parameters $r_n$ and $x_n^c$, we firstly construct the histogram with number of bins much smaller than the number of samples. The parameter $r_n$ is the relative frequency and $x_n^c$ is the center of the $n$-th bin. The values of $\delta_n$ depends on the real PDF and, therefore, it is the variable to be optmized. The number of neurons in the hidden layer is equal to the number of bins of the histogram. The first test was realized with a normal distribution

$$p(x) = \frac{1}{\sqrt{2\pi}} e^{-\frac{x^2}{2}}. \tag{11}$$

Twenty thousand numbers were randomly chosen according to the PDF in eq. (11) and a histogram with 2000 bins was constructed, hence, there were two thousand neurons in the hidden layer. The real and estimated PDF's can be seen in Fig. 5. The value of $\delta_n$ used was 30 for all neurons in the hidden layer.

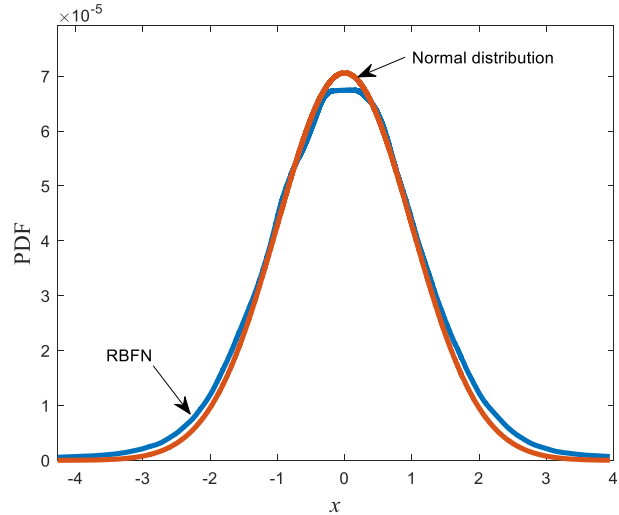

Fig. 5 – Normal distribution and its estimation using a RBFN with $W_2$.

The second test uses a standard Cauchy distribution:

$$p(x) = \frac{1}{\pi(1+x^2)}. \tag{12}$$

Ten thousand numbers were randomly chosen according to the PDF in eq. (12) and a histogram with 2000 bins was constructed. The real and estimated PDF's can be seen in Fig. 6. The value of $\delta_n$ used was 0.5 for all neurons in the hidden layer.

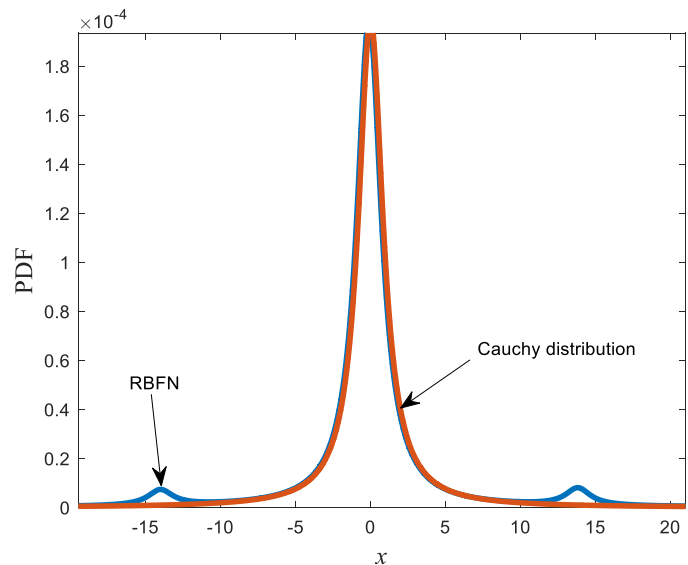

Fig. 6 – Cauchy distribution and its estimation using a RBFN with $W_2$.

As one can note, in both cases the PDF estimation was not perfect but it was good enough for, for example, to provide a good estimation of the entropy or disentropy.

## 5. Conclusions

The good performances of the two RBFNs presented show that the proposed kernel based on the Lambert-Tsallis $W_q$ function, eq. (2), is usefull and it can be used in RBFNs. Here, we were restricted to $q = 2$ because of its easy and fast numerical calculation, but other values of $q$ can also be tested. At last, the high rate success of the classifier also shows that the quantum relative disentropy is a valid measure of distance between quantum states.

## Acknowledgements

This study was financed in part by the Coordenação de Aperfeiçoamento de Pessoal de Nível Superior - Brasil (CAPES) - Finance Code 001, and CNPq via Grant no. 307184/2018-8. Also, this work was performed as part of the Brazilian National Institute of Science and Technology for Quantum Information.